# Real-Time Contingency Analysis with Corrective Transmission Switching —Part II: Results and Discussion

Xingpeng Li, *Student Member, IEEE*, Mostafa Sahraei-Ardakani, *Member, IEEE*, Pranavamoorthy Balasubramanian, *Student Member,* Mojdeh Abdi-Khorsand, *Student Member, IEEE*, Kory W. Hedman, *Member, IEEE*, and Robin Podmore, *Fellow, IEEE*

*Abstract*—This paper presents the performance of an AC transmission switching (TS) based real-time contingency analysis (RTCA) tool that is introduced in Part I of this paper. The approach quickly proposes high quality corrective switching actions for relief of potential post-contingency network violations. The approach is confirmed by testing it on actual EMS snapshots of two large-scale systems, the Electric Reliability Council of Texas (ERCOT) and the Pennsylvania New Jersey Maryland (PJM) Interconnection; the approach is also tested on data provided by the Tennessee Valley Authority (TVA). The results show that the tool effectively reduces post-contingency violations. Fast heuristics are used along with parallel computing to reduce the computational difficulty of the problem. The tool is able to handle the PJM system in about five minutes with a standard desktop computer. Time-domain simulations are performed to check system stability with corrective transmission switching (CTS). In conclusion, the paper shows that corrective switching is ripe for industry adoption. CTS can provide significant reliability benefits that can be translated into significant cost savings.

## I. INTRODUCTION

PREVIOUS research has demonstrated that transmission switching (TS) offers a variety of benefits. Despite the vast body of literature that is dedicated to TS, industry adoption has been very limited. The barriers include the following: 1) TS problems are computationally expensive. 2) Studies on large-scale real systems based on actual operations is very rare and, thus, verifiable results are rare. 3) Many studies rely on many algorithmic approximations, e.g., DC power flow assumptions. 4) System stability is also a concern. For a more extensive literature review, refer to part I [1] of this paper. This two-part paper aims to address these concerns.

The above-mentioned state of the art challenges are investigated with the proposed fast TS-based AC real-time contingency analysis (RTCA) package, as described by Part I of this two-part paper. Stability studies are performed on a subset of the CTS solutions to confirm system stability. While Part I discusses the methodology, Part II includes description of the data, results, and discussion.

Actual snapshots from energy management systems (EMS) obtained from the Electric Reliability Council of Texas (ERCOT) (3 snapshots) and the Pennsylvania New Jersey Maryland (PJM) Interconnection (167 snapshots) are used as the inputs to the corrective transmission switching tool; furthermore, the Tennessee Valley Authority (TVA) provided data corresponding to three days, which were used to produce 72 base case AC power flows. The results confirm the effectiveness of the developed tool, which can readily be adopted by the industry. To our knowledge, this paper is among the first comprehensive studies that addresses the state of the art challenges of TS with actual EMS data at this level of detail.

Over 1.5 million contingencies are simulated on the data from TVA, ERCOT, and PJM to analyze the effectiveness of corrective transmission switching (CTS) with RTCA. The results show that 10%-33% of the contingencies with post-contingency violations would have no violations if a single post-contingency corrective transmission switching action is implemented. Substantial reductions in post-contingency violations are observed on 56%-83% of the cases. The solution time to achieve such results is reasonable for real-time implementation. The computational efficiency is attained by using fast heuristics, explained in Part I, as well as parallel computing. Overall, the results are very promising and suggest that CTS is ripe for industry adoption for the RTCA application.

The rest of this paper is organized as follows: Section II describes the EMS data received and presents vanilla contingency analysis results. Regular contingency analysis without CTS is referred to as "vanilla contingency analysis." Section III presents the results obtained from the RTCA CTS routine for these three systems. A comprehensive discussion of the results is also presented. Section IV presents the computational time and the speed up achieved from parallel computing. This section sheds light on high performance computing (HPC) for the proposed method. Section V presents the stability analysis results performed on the PJM system for selected cases and provides insight into dynamic stability of CTS. Finally, Section VI concludes this paper.

## II. EMS DATA AND VANILLA CONTINGENCY ANALYSIS

Data is obtained from three reliability coordinators (RC): TVA, ERCOT, and PJM. The characteristics of the data are summarized in Table I.

The research is funded by Department of Energy (DOE) Advanced Projects Agency – Energy (ARPA-E) under the Green Electricity Network Integration (GENI) program.

X. Li, M. Sahraei-Ardakani, P. Balasubramanian, M. Abdi-Khorsand, and K. W. Hedman are with the School of Electrical, Computer, and Energy Engineering, Arizona State University, Tempe, AZ, 85287, USA (e-mail: {Xingpeng.li; Mostafa; pbalasu3, mabdikho}@asu.edu; kwh@myuw.net).

R. Podmore is the founder and president of IncSys, Bellevue, WA, 98007, USA, (e-mail: robin@incsys.com).

TABLE I
DESCRIPTION OF THE ACTUAL SYSTEM DATA

| System | Scenarios | Load (Real GW, Reactive GVAr) | Buses | Generators | Branches |
|---|---|---|---|---|---|
| TVA | 72 | ~(24, 4) | ~1.8k | ~350 | ~2.3k |
| ERCOT | 3 | ~(57, 8) | ~6.4k | ~700 | ~7.8k |
| PJM | 167 | ~(139,22) | ~15.5k | ~2,800 | ~20.5k |

Load profiles for 72 hours were obtained from TVA along with TVA's network information. Detailed information on TVA can be found in [2]. Security-constrained unit commitment (SCUC), which includes a DC optimal power flow, was run on the data to obtain 72 operating points for TVA. This SCUC solution was then used as a starting solution to obtain AC power flow base case solutions. If network violations are observed in the base case, out of market corrections [3] are made to obtain AC feasibility. This AC solution is the basis of the analysis for the TVA system conducted in this paper. This approach was taken based on the available data from TVA.

The EMS data from ERCOT and PJM is directly used and all of the analysis is done on the original EMS snapshots with no modifications. EMS data for 167 hours, which correspond to a week in July 2013, was provided by PJM. ERCOT provided three snapshots of EMS data; these hours correspond to critical winter storms that led to operation difficulties.

Contingency analysis is run on all three systems to identify the contingencies that would result in network violations. Voltage violations are recorded for values outside the range of 0.9 p.u. to 1.1 p.u. Transmission flow violations occur when the flow exceeds the emergency ratings. The threshold for significance of voltage violations is assumed to be 0.005 p.u. and the threshold for thermal flow violation is set at 5 MVA, both on a system aggregate level. Violations less than these thresholds are ignored due to their insignificance. Buses and transmission assets below 70 kV are not monitored. This is consistent with existing practices in industry.

Table II summarizes the results of this initial vanilla contingency analysis. The table shows that the original dispatch is vulnerable to a number of contingencies for all three systems. A full $N$-1 study is conducted and all contingencies with violations (beyond the specified threshold) are sent to the CTS routine. Table II shows that the percentage of contingencies with violations for TVA is larger than ERCOT and PJM. Moreover, the percentage of contingencies for which the violations are within the tolerance for TVA is also notably greater than ERCOT and PJM. The reason for such differences is that the TVA AC power flow base cases were created based on the data provided by TVA whereas ERCOT and PJM data came directly from their EMS systems. In real-time operations, the system operators perform adjustments that would make the operations less vulnerable to contingencies. Thus, there would naturally be a significant difference between the ERCOT and PJM actual EMS cases and the TVA cases that were created since the TVA data did not go through such a process.

It should be noted that system operators have ways to handle some of these contingencies via special protection schemes (SPS) [4], flexible AC transmission system (FACTS) devices [5]-[7], switchable shunts [8], transformer tap setting adjustment [8], or other corrective mechanisms. While such other preventive or corrective approaches can also be used instead of corrective transmission switching, the results clearly demonstrate the breadth and depth to which corrective transmission switching can benefit system operations. This approach identifies CTS solutions in real-time, unlike offline techniques that are not guaranteed to work for all operating states.

TABLE II
OVERALL STATISTICS ON RTCA SIMULATIONS

| System | # of Contingencies Simulated | # of Contingencies with Violations | # of Contingencies with Violations beyond Threshold |
|---|---|---|---|
| TVA | 126,449 | 15,540 | 4,272 |
| ERCOT | 13,044 | 52 | 40 |
| PJM | 1,437,749 | 11,100 | 8,064 |

III. CASE STUDIES

Different CTS strategies, discussed in Part I, are implemented and the benefits obtained from each methodology are analyzed. In summary, the following heuristics are used: 1) the 100 closest branches (transmission lines or transformers) to the contingency element (CBCE), 2) 100 closest branches to the violation element (CBVE), and 3) data mining (DM).

Table III presents the overall statistics on the CTS simulations. All the results presented in Table III correspond to the benefits obtained from the first best switching action as identified from the CBVE proximity search algorithm. A beneficial CTS solution may reduce the aggregate network violations without ensuring a Pareto improvement (PI); however, note that it is easy to select CTS solutions that only provide Pareto improvements and the difference between enforcing a PI solution or not produce very similar results.

Table IV presents the average violation reduction with CTS as an average percentage. The metric is defined in Part I of this paper. The average thermal flow violation reductions are 40%, 53%, and 59% for TVA, ERCOT, and PJM respectively. Similarly, the voltage violation reductions on average are found to be 36%, 12%, and 20% for TVA, ERCOT, and PJM, respectively. Table IV shows that the violation reductions with and without consideration of Pareto improvement are not very different. This finding illustrates that the CTS actions identified in response to a specific violation almost never induces additional violations in the system. This is an important and interesting finding supported by evidence shown in Table IV.

TABLE III
OVERALL STATISTICS ON RTCA CTS SIMULATIONS

| System | # of Contingencies Fully Eliminated | # of Contingencies with Partial Viol. Reduction | # of Contingencies with No Viol Reduction |
|---|---|---|---|
| TVA | 427 (6 per hour) | 3,535 (49 per hour) | 310 (4 per hours) |
| ERCOT | 6 (2 per hour) | 27 (9 per hour) | 7 (2 per hour) |
| PJM | 2,684 (16 per hour) | 4,554 (27 per hour) | 826 (5 per hour) |

TABLE IV
AVERAGE VIOLATION REDUCTION

| System | Avg. Flow Violation Reduction | | Avg. Voltage Violation Reduction | |
|---|---|---|---|---|
|  | w/o PI | w/ PI | w/o PI | w/ PI |
| TVA | 40.0% | 40.0% | 36.2% | 35.6% |
| ERCOT | 53.1% | 49.3% | 12.3% | 12.3% |
| PJM | 59.3% | 59.0% | 19.5% | 19.3% |

*A. TVA System*

For the TVA system, all heuristics, CBCE, CBVE, and DM, are implemented. Three DM approaches are constructed in a way that the candidate list for switching actions in a particular day will determine the beneficial CTS solutions identified for the other two days. Additional details on the different methodologies used are available in Part I of the paper.

Table V presents the results obtained from these CTS heuristics. Even though it is expected that the CBCE approach would perform similar to the CBVE, the reduction in violations obtained with both methods is found to be different for the TVA system. The majority of critical contingencies are generator contingencies for the TVA system, which involves generation redispatch from units spread across the system. With the redispatch, violations may not be close to the initial contingency. Hence, the effect of switching lines in the proximity of a contingency is different from the effects of switching a line in the proximity of a line that is overloaded. The results from complete enumeration (CE) are given to show the effectiveness of the different heuristics. The CBVE approach provides 40% reduction in thermal flow violations in comparison with 40.8% reduction achieved by CE. However, the reduction in voltage violation with CBVE method is only 36.2% compared to 48.2% that is achieved with CE. CBVE takes only 6.8% of the time that CE takes; the results show that CBVE is fast and accurate. The data mining approach performs the best amongst the heuristics. All data mining methods provide similar violation reductions. The solution time for DM3 is significantly smaller. DM3 provides 26 times faster solutions with almost the same accuracy in comparison to CE. DM3 chooses the fewest candidate lines for its list, which is why it is the fastest. The solution times in Table V is with a single processor and does not involve parallel processing.

TABLE V
RESULTS FROM VARIOUS CTS METHODS ON THE TVA SYSTEM W/O HPC

| TS Method | Avg. Solution time (s) | Avg. Flow Violation Reduction | | Avg. Voltage Violation Reduction | |
|---|---|---|---|---|---|
| | | w/o PI | w/ PI | w/o PI | w/ PI |
| CBCE | 167 | 15.6% | 15.0% | 31.8% | 30.9% |
| CBVE | 178 | 40.0% | 40.0% | 36.2% | 35.6% |
| DM1 | 202 | 40.6% | 40.1% | 48.1% | 47.8% |
| DM2 | 107 | 40.5% | 40.0% | 48.1% | 47.7% |
| DM3 | 98 | 40.5% | 40.0% | 48.0% | 47.7% |
| CE | 2585 | 40.8% | 40.3% | 48.2% | 47.9% |

Table VI presents the solution time for the original RTCA as well as the various CTS heuristics implemented on the TVA system. Note that the solution time reported for CTS is averaged over all 72 hours and does not include the solution time required for performing the original RTCA. In order to be consistent, the solution time is reported in the same way through the remainder of this paper. It is found that the solution time for CBVE method is about 4 times longer than the solution time required for performing RTCA; however, the DM3 method requires only twice the time that is required for performing RTCA. It is important to note that the maximum solution time to identify such quality CTS solutions is less than 4 minutes even with sequential processing on a computer with moderate computing capability.

Fig. 1 shows both flow violation reductions and voltage violation reductions associated to the five best CBVE switching actions, without enforcing a Pareto improvement. The average depth of the five best candidates is around 40 in the candidate list. It is found that the reduction in violation obtained with and without enforcing the solution to be a Pareto improvement is very similar for any of the approaches tested. This implies that the CTS solutions rarely cause additional violations while trying to reduce the original post-contingency violations. The figure shows that, as the rank of the switching candidate increases, the thermal flow violation reduction drastically falls; however, the variation in voltage violation reduction is not so steep. It should be noted that these results are specific to the TVA system that is used for the analysis and a generalization cannot be made based on these results for other systems. The magnitude of congestion, as one of the determinants of the effectiveness of this technology, is drastically different from one system to another. Other factors such as reserve requirements, type of generators, and the topology of the network also play important roles in performance of CTS. Moreover, this analysis is conducted on the data corresponding to 3 days in September 2012; such results will vary throughout the year.

TABLE VI
SOLUTION TIME FOR RTCA WITH CTS ON THE TVA SYSTEM W/O HPC

| TS Method | Average (s) | Min (s) | Max (s) |
|---|---|---|---|
| RTCA | 45 | 43 | 48 |
| CBCE | 167 | 17 | 346 |
| CBVE | 178 | 18 | 373 |
| DM1 | 202 | 18 | 464 |
| DM2 | 107 | 10 | 231 |
| DM3 | 98 | 10 | 207 |
| CE | 2585 | 209 | 10524 |

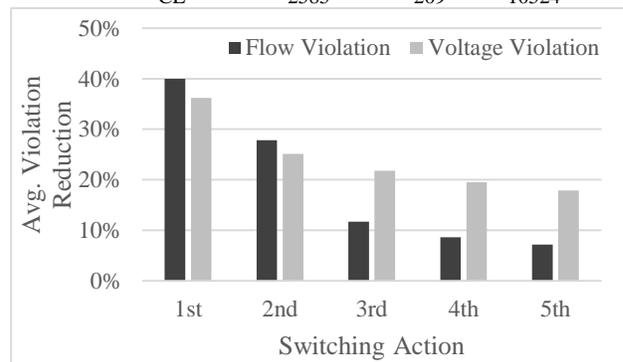

Fig. 1. Violation reductions with CTS actions on the TVA system.

*B. ERCOT System*

ERCOT provided three snapshots of their system. Two different heuristics, CBVE and CBCE, are used to identify the corrective switching actions to reduce the post-contingency violations. Complete enumeration of all the switching actions is also performed to find the upper bound of the benefits that can be obtained with CTS. Due to limited available data, data mining is not performed on the ERCOT system. Table VII lists the results of various transmission switching methods on the ERCOT system. It is found that both CBVE and CBCE methods provide similar benefits in terms of the reduction in voltage violations. However, CBVE results in 10% more reductions in thermal flow violations. The reduction in violations achieved with both CBVE and CBCE heuristics are very similar to that achieved through complete enumeration, which confirms the accuracy of the heuristics. Note that both heuristics achieved such quality solutions 47 times faster than CE.

TABLE VII
RESULTS OF VARIOUS CTS METHODS ON ERCOT SYSTEM W/O HPC

| TS methods | Avg. Solution time (s) | Avg. Flow Violation Reduction | | Avg. Voltage Violation Reduction | |
|---|---|---|---|---|---|
| | | w/o PI | w/ PI | w/o PI | w/ PI |
| CBCE | 245 | 40.8% | 37.7% | 12.1% | 12.1% |
| CBVE | 244 | 53.1% | 49.3% | 12.3% | 12.3% |
| CE | 11,505 | 53.3% | 49.3% | 14.3% | 14.3% |

Table VIII presents the average, minimum, and maximum solution times for RTCA and for the CTS heuristics. The overall solution times for the CTS heuristics are found to be less than the time taken for RTCA since the number of critical contingencies that require CTS is smaller for the ERCOT system compared to TVA. The maximum solution time to find the CTS actions is close to 6 minutes even for sequential implementation of the CTS heuristics.

Fig. 2 presents the thermal flow and voltage violation reductions corresponding to the top 5 switching actions obtained through the CBVE heuristic. Note that, even the fifth candidate provides significant flow violation reductions. It is found that, for the ERCOT system, significant reductions in thermal flow violation is achieved by all the three methods; however, the voltage violation reduction is comparatively smaller.

TABLE VIII
SOLUTION TIME OF RTCA AND VARIOUS TRANSMISSION SWITCHING METHODS ON ERCOT W/O HPC

| TS Method | Average (s) | Min (s) | Max (s) |
|---|---|---|---|
| RTCA | 767 | 575 | 785 |
| CBCE | 245 | 182 | 356 |
| CBVE | 244 | 185 | 350 |
| CE | 11505 | 8728 | 16734 |

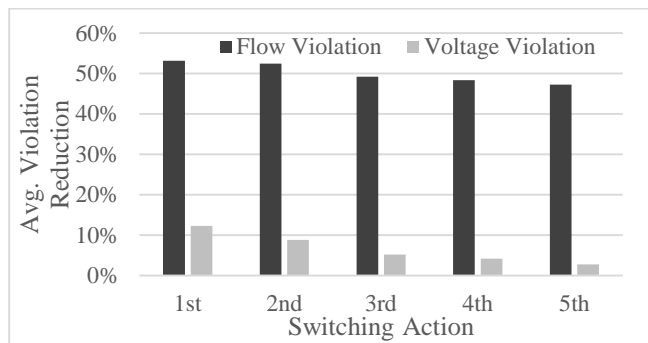

Fig. 2. Violation reductions with CTS actions on the ERCOT system.

Fig. 3 presents an actual example from the ERCOT system, which illustrates how CTS eliminates transmission flow violations. Fig. 3 (a) shows the normal operating condition. There are two power plants in the subsystem as shown in the figure. Fig. 3 (b) shows the post-contingency state after the loss of a major transmission line, which exports 1 p.u. of flow to the rest of the network in normal condition. The loss of this exporting line results in overloading of two lines due to the excess injection of power in the subsystem. One way to resolve the problem would be switching off an importing line to reduce the injection of the power into the circuit. Fig. 3 (c) shows the system after switching a transmission line that imports 2.3 p.u. into this subsystem. This switching action effectively eliminates the violations by reducing the flow on the overloaded lines and rerouting power through other paths in the network. Note that all numbers are masked in a per unit basis to protect proprietary data.

## C. PJM System

The PJM system is the largest of the three systems used for the analysis. Hence, the computational time to run CTS-based RTCA on PJM is significantly longer compared to TVA and ERCOT. Therefore, all simulations on the PJM system are performed using HPC. For this specific section, Section III.C, 6 threads are only used. The simulation is performed on the same machine that was used for TVA and ERCOT system, with the one exception that ERCOT and TVA were solved sequentially with only 1 thread at a time.

Similar to TVA and ERCOT, a vanilla contingency analysis is performed first to identify contingencies that would lead to violations. Similar to the ERCOT system, the two CTS heuristics, CBCE and CBVE, are used to form a rank list consisting of potential switching candidates for the PJM system.

Table IX presents the statistics for violation reductions corresponding to the 5 best switching solutions with CBVE heuristic. The percentage reduction in flow violations is found to be 59% and 46% for the first and the fifth best CTS actions respectively. However, in case of voltage violation reductions, it varies from 20% to 6% for the first and the fifth best switching actions. Note that the depth is relatively small and increases as the solutions become less beneficial, which demonstrates the effectiveness of the proposed heuristic methods. The results emphasize that quality solutions are found within the close vicinity of the elements with violations.

Table X presents the flow violation reduction and voltage violation reduction for the corresponding top five switching candidates with CBCE heuristic. It is found that all the top 5 CTS solutions provide significant reductions in thermal flow violation for the PJM system; however, only the top 3 CTS solutions provide voltage violation reductions above 10%. Table X also presents the statistics for distance as defined in part I of this two-part paper. The average distance of the identified CTS solutions to the contingency element is around 1-2 for flow violations and about 3 for voltage violations.

The detailed solution time for RTCA and the two CTS heuristics are presented in Table XI. The CTS results presented in Tables IX, X, and XI show that both heuristics perform equally well with respect to flow violation reductions, voltage violation reductions, and solution time on the PJM system.

TABLE IX
RESULTS OF THE 5 BEST SWITCHING ACTIONS ON PJM SYSTEM WITH CBVE

| Candidate | Flow Violations | | | | Voltage Violations | | | |
|---|---|---|---|---|---|---|---|---|
| | w/o PI | | w/ PI | | w/o PI | | w/ PI | |
| | Avg. Reduc. | Depth | Avg. Reduc. | Depth | Avg. Reduc. | Depth | Avg. Reduc. | Depth |
| 1st Best | 59% | 14.9 | 59% | 15.4 | 20% | 37.8 | 19% | 38.6 |
| 2nd Best | 58% | 17.4 | 57% | 17.7 | 15% | 38.8 | 14% | 39.1 |
| 3rd Best | 53% | 23.9 | 52% | 24.7 | 12% | 38.2 | 11% | 38.9 |
| 4th Best | 49% | 27.5 | 49% | 26.9 | 8% | 40.9 | 8% | 40.7 |
| 5th Best | 46% | 28.1 | 46% | 28.4 | 6% | 42.2 | 6% | 42.4 |

TABLE X
STATISTICS OF THE 5 BEST SWITCHING ACTIONS ON THE PJM SYSTEM WITH CBCE HEURISTIC W/O PI

| Candidate | Avg. for Flow Violation | | | Avg. for Voltage Violation | | |
|---|---|---|---|---|---|---|
| | Reduction | Depth | Distance | Reduction | Depth | Distance |
| 1st Best | 61.6% | 18.2 | 1.36 | 19.1% | 39.8 | 3.16 |
| 2nd Best | 58.1% | 21.9 | 1.55 | 14.2% | 40.4 | 3.24 |
| 3rd Best | 55.6% | 26.6 | 1.90 | 10.9% | 40.1 | 3.21 |
| 4th Best | 49.3% | 31.3 | 2.15 | 7.2% | 41.6 | 3.34 |
| 5th Best | 45.3% | 31.3 | 2.15 | 5.9% | 41.7 | 3.30 |

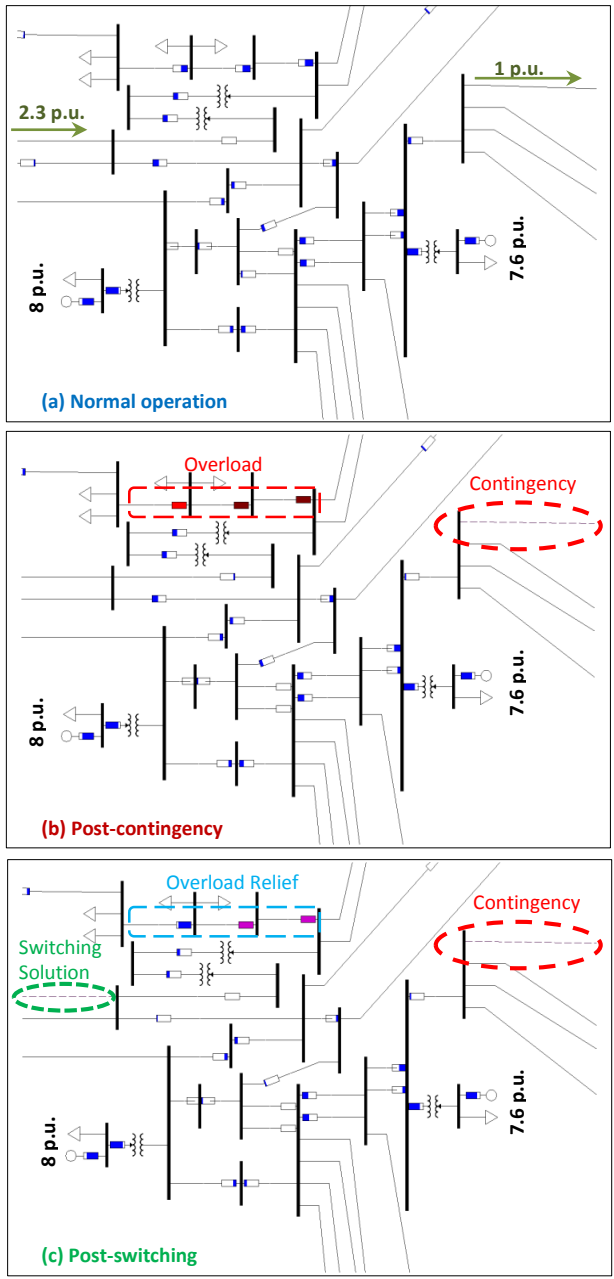

Fig. 3. An actual example from ERCOT showing how CTS eliminates flow violations. (a): normal operation; (b): post-contingency; (c): post-switching.

TABLE XI
SOLUTION TIME OF RTCA AND VARIOUS TRANSMISSION SWITCHING METHODS ON PJM SYSTEM W/HPC (6 THREADS)

| TS Method | Average (s) | Min (s) | Max (s) |
|---|---|---|---|
| RTCA | 2617 | 2187 | 3100 |
| CBCE | 1593 | 237 | 3499 |
| CBVE | 1612 | 242 | 3441 |

In order to estimate the quality of solutions obtained from the two CTS heuristics, complete enumeration of all possible switching actions is performed on 6 selected EMS snapshots. The hours represent sample data for peak, off-peak, and shoulder hours. Table XII presents the violation reductions and the corresponding computational time for the complete enumeration method as well as CBCE and CBVE heuristics. The results show that both the heuristic methods perform close to complete enumeration. The significant advantage of the heuristics is that the solution time to achieve such good quality CTS actions is 110 times faster than complete enumeration.

The results presented in Table XII confirm the effectiveness of the two heuristics to find quality solutions quickly. Furthermore, almost all of the CTS solutions make Pareto improvements and no significant difference was observed between the heuristics and CE in this sense.

Fig. 4 shows how the five candidates perform in one particular contingency case. This contingency resulted in the overload of only a single line in the system. The best switching action provided a 100% reduction in violation, while the fifth best CTS action provided 18% reduction in violation. All five switching actions provide Pareto improvements. Fig. 5 presents an artificially created example that conceptually shows the case discussed in Fig. 4. There is power flow from bus 1 towards buses 6, 7, 10 and the rest of the system as seen in Fig. 5 (a). A contingency on line connecting buses 4 and 6 creates a flow violation on the parallel path connecting buses 4 and 5 as shown in Fig. 5 (b). The top 5 switching actions identified by the CTS tool and the corresponding flow violation reductions for the first two CTS solutions on the overloaded line are presented in Fig. 5 (c) and (d) respectively. Note that the percentage loading on the lines presented in Fig. 5 (a) is based on the normal rating, 'RATE A' and the percentage loading in the rest of the post contingency cases are presented with respect to the emergency rating, 'RATE C'.

In another instance, a particular contingency caused an aggregate voltage violation of 0.4 pu spread across 17 buses. All of the top five switching actions fully eliminate the violations.

As described above, an important observation made for all the test cases is that the results with and without Pareto improvement are very similar, suggesting that the majority of the switching solutions provide Pareto improvement automatically. Note that, for each contingency, only the five best switching candidates are proposed to the operator. Non-Pareto solutions are very unlikely to be among those best solutions.

TABLE XII
RESULTS OF VARIOUS CTS METHODS ON PJM SYSTEM FOR SELECT HOURS

| TS Method | Avg. Solution Time (s) | Avg. Flow Violation Reduction | | Avg. Voltage Violation Reduction | |
|---|---|---|---|---|---|
| | | w/o PI | w/ PI | w/o PI | w/ PI |
| CBCE | 872 | 62.1% | 61.0% | 19.4% | 19.4% |
| CBVE | 875 | 59.4% | 59.4% | 19.4% | 19.4% |
| CE | 96922 | 62.5% | 62.5% | 21.0% | 20.4% |

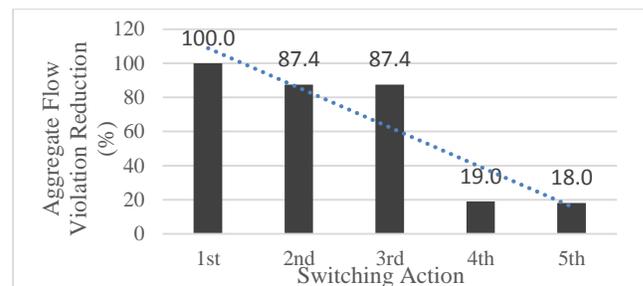

Fig. 4. Reduction in worst case flow violation corresponding to top 5 CTS actions on the PJM system.

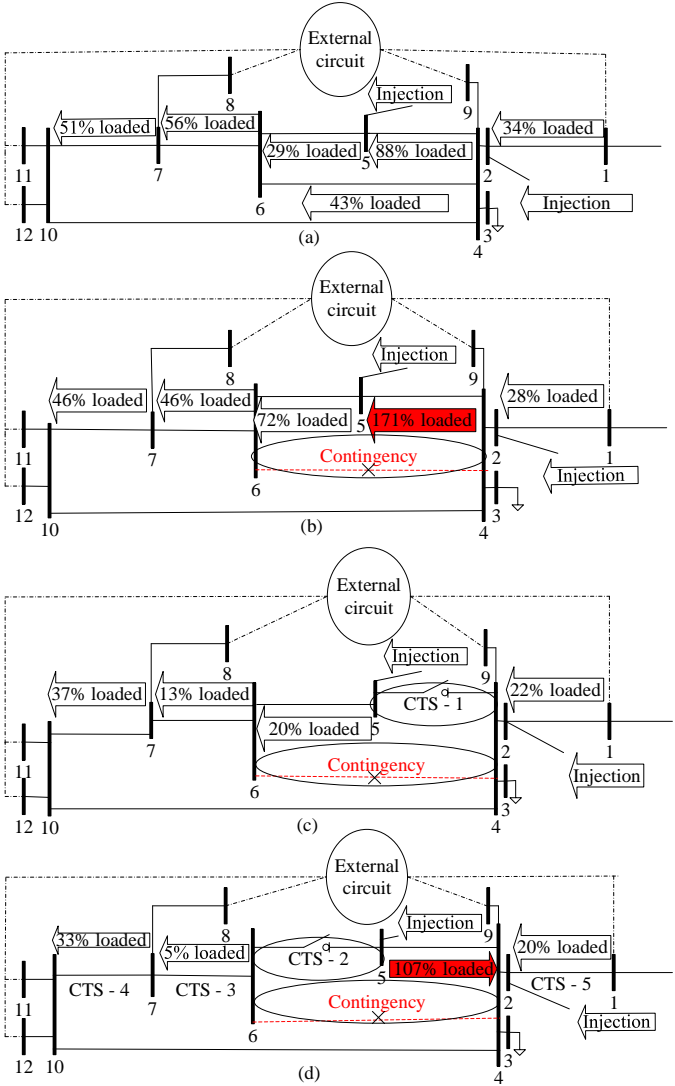

Fig. 5. An artificially created example that conceptually represents the worst case flow violation. The performance of the top five CTS actions on the PJM system is shown: (a) Pre-contingency case, (b) Post-contingency case, (c) Post switching – candidate 1, (d) Post switching – candidate 2. The other candidates 3-5 are also shown, which resulted in violations of 7%, 57% and 58% respectively.

## IV. HIGH PERFORMANCE COMPUTING

Computational complexity of transmission switching has been one of the factors inhibiting the application of optimization-based approaches for transmission switching. With the use of heuristics, the computational time presented in the previous section is substantially reduced for RTCA. Furthermore, with the advancements in technology, parallel computing can improve the computational efficiency of the problem. Given the nature of the problem, which includes RTCA and testing the list of switching candidates, computational time is expected to drastically improve with parallel computing. The speedup is investigated for all the three systems with parallel computing. The high performance clusters from Lawrence Livermore National Laboratory are used for this analysis.

Table XIII presents the average solution time in seconds with different number of threads for vanilla contingency analysis and CTS on the three large-scale systems. It is observed that as the number of threads increased, the solution time decreases as expected. Up to 128 threads were used for vanilla contingency analysis. The solution time for RTCA on the TVA system comes down to 0.7s as compared to 49s for a sequential run with a single thread. The RTCA solution time for the ERCOT reduced to 10s from around 900s without parallel processing. For PJM, the solution time with 8 threads in parallel is almost half an hour and it decreases to about two minutes with 128 threads in parallel. The parallel efficiency of vanilla contingency analysis for the PJM system is presented in Fig. 6. This metric is presented and explained in (3) in Part I of this paper. Since the candidate list length for both CBVE and CBCE methods was chosen to be 100 elements, using more than 100 threads for CTS will not be beneficial, unless the power flow algorithm itself is parallelized. The variation in solution time for CTS with CBVE heuristic, for different number of threads, is presented in Table XIII. The results show that a computer cluster with only 100 processors can handle a snapshot of PJM data in less than two minutes on average.

Note that all the $N$-1 events are simulated for the analysis associated with Table XIII. System operators usually run their contingency analysis on a critical contingency list, which is a subset of all the $N$-1 contingencies. Thus, the computational time presented in Table XIII is expected to be reduced for actual implementation. The CTS time for PJM can be further reduced to well below a minute if the critical contingency list is available and only those contingencies are modeled.

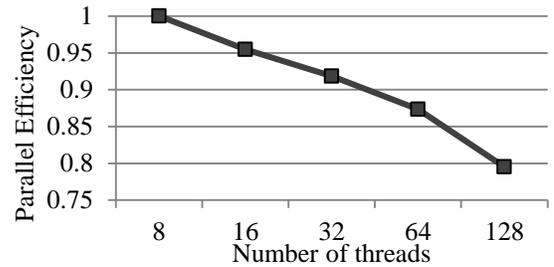

Fig. 6. Parallel efficiency of vanilla contingency analysis for the PJM system.

TABLE XIII
AVERAGE SOLUTION TIME FOR RTCA AND CTS WITH DIFFERENT THREADS

| | RTCA | | | | CTS | | |
|---|---|---|---|---|---|---|---|
| # of Threads | TVA | ERCOT | PJM | # of Threads | TVA | ERCOT | PJM |
| 1 | 49 | 899 | NA | 1 | 172 | 280 | NA |
| 2 | 25 | 455 | NA | 2 | 89 | 142 | NA |
| 4 | 13 | 231 | NA | 4 | 47 | 74 | NA |
| 8 | 6.9 | 123 | 1634 | 8 | 27 | 41 | 999 |
| 16 | 3.7 | 63 | 856 | 16 | 16 | 23 | 566 |
| 32 | 2.0 | 33 | 445 | 25 | 11 | 15 | 323 |
| 64 | 1.1 | 18 | 234 | 50 | 7.2 | 8.4 | 173 |
| 128 | 0.7 | 10 | 128 | 100 | 6.6 | 5.6 | 96 |

## V. STABILITY STUDIES ON THE PJM SYSTEM

The overall results obtained from the CTS heuristics are effective and efficient with respect to the quality of solutions and the solution time. The algorithms scale well for large-scale systems and all the analysis is done on an AC framework. However, the stability of the system following a switching action needs to be investigated.

NERC requires system operators to have plans for loss of a second bulk element following an initial contingency ($N$-1-1).

NERC specifically requires system operators to maintain dynamic stability following an *N*-1-1 event [9]. A CTS action following a contingency can be seen as an *N*-1-1 event, as a second element is being taken out of the system. Therefore, according to the NERC standard, it is required that the system maintains dynamic stability following a CTS action. An unstable corrective switching action, thus, would show that the NERC requirements are being violated, which would require attention in and of itself without even considering the technology of corrective transmission switching. Furthermore, note that the corrective switching action is on a line that does not have a fault current; since systems should be protecting against two (*N*-1-1) faults, the corrective switching action should not cause a stability concern. After in-person discussions with PJM, MISO, ERCOT, and ISONE, they also confirm that stability for a single post-contingency switching action (after the system regains steady-state after the contingency) should not be a major hindrance to such a technology, which is one reason why PJM already implements this technology today, based on offline analysis [10]; moreover, PJM has confirmed that the decision to run a stability study before implementing a switching solution is at the discretion of the operator [11]. While these arguments in support of corrective transmission switching not being a primary concern for stability, nonetheless, it is important to analyze the impact on stability. Therefore, to test this hypothesis and to check for dynamic security, a subset of switching solutions are tested for stability.

The stability studies are conducted on specific hours of the system spreading across the entire week of the PJM data. The specific hours for testing the stability of the CTS actions were chosen based on different loading conditions and the number of critical contingencies present for that particular hour. Samples of peak, off peak, and shoulder hours are chosen along with the hour that have the maximum number of critical branch contingencies and the hour that has the maximum number of critical generator contingencies. Overall, the stability analysis is performed on 5 EMS snapshots with completely different system operating states.

Time domain simulations are performed on all contingencies that have violations for the selected hours. Two different methodologies are followed to perform the time domain simulation for the branch contingencies and the generator contingencies. The detailed methodology is presented in Part I of this paper. Stability analysis is conducted to examine if the proposed corrective transmission switching solutions cause instability; in total, 284 contingencies, along with the CTS solutions, are analyzed. Overall, only 2 (0.7%) of the cases that were tested failed transient stability analysis. Fig. 7 presents the time domain simulation response for a branch contingency with CTS to relieve voltage violations in the system. Note that this particular contingency resulted in voltage violations on 17 buses with an aggregate violation of 0.4 pu. The CTS action completely eliminates those voltage violations. Fig. 8 represents the time domain simulation for a generator contingency that caused thermal flow violations. This particular contingency resulted in the maximum flow violation among all generator contingencies tested. The CTS action eliminates the flow violations completely.

Overall, more than 99% of the top switching candidates provide a stable solution, which is expected according to NERC standards. Note that only 0.7% of the cases, which were tested, have a transient rotor angle stability issue associated with the switching action. These results are expected as PJM is reported to have limited concerns regarding transient stability for their system [12].

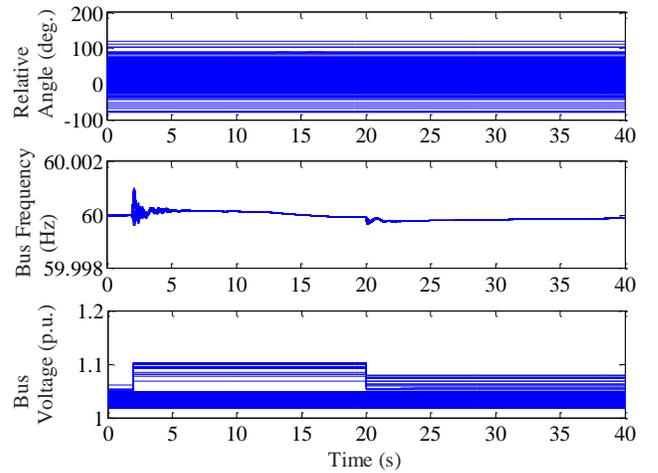

Fig. 7. Time domain simulation for a transmission contingency with CTS action on a lightly loaded hour on the PJM system.

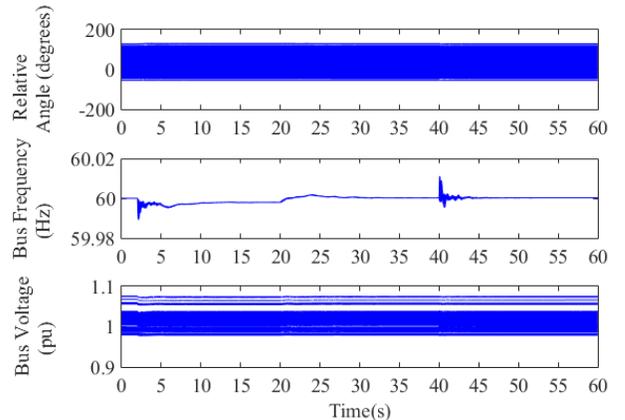

Fig. 8. Time domain simulation for a generator contingency with generation redispatch and CTS.

## VI. CONCLUSIONS

Corrective transmission switching provides the operators with an alternative tool to handle potential post-contingency violations. The tool developed in this paper takes in operational data of real large-scale power systems (PJM, ERCOT, and TVA) in PSS/E .raw format. Simulation results validate the effectiveness of CTS as significant post-contingency violation reductions are achieved with CTS.

The computational efficiency is achieved by using extremely powerful heuristics along with parallel computing. The package is AC based and there is no loss of precision as often observed in DC-based algorithms. The simple heuristics used for the CTS routine are able to find quality solutions very quickly. Local search algorithms around the contingency may not perform as well for generator contingencies due to the spatial distribution of redispatch and the resulting violations. Data mining methods may not perform well if the system condition changes significantly from the training data. Overall, the dynamic search around the violations shows the most promising performance. Parallel computing is effectively used for further reduction of the computational time. Moreover, stability analysis is performed to check the stability of the proposed corrective TS actions. A subset of CTS actions is analyzed and the results show that more than 99% of the CTS actions do not cause instability issues.

To conclude, the proposed technology is able to very quickly propose quality CTS solutions to alleviate post-contingency violations. The promising results presented in this paper show that transmission switching is ripe for industry adoption for the real-time contingency analysis application.

ACKNOWLEDGEMENT

The authors would like to thank Dr. Deepak Rajan for his support with HPC implementation and Lawrence Livermore National Lab. Discussions with Dr. Vijay Vittal on stability issues, as well as feedback from Chris Mosier for the RTCA tool development, are greatly appreciated. The authors would also like to express thanks for valuable feedback from PJM, ERCOT, TVA, ISONE, MISO, Alstom Grid, and ABB.

OPEN SOURCE SOFTWARE ACCESS

The proposed technology was built around IncSys' and PowerData's open source decoupled power flow; interested parties can download the software [13]. For the proposed real-time contingency analysis with corrective transmission switching, which is implemented with MPI based HPC, interested parties can email Dr. Kory W. Hedman (kwh@myuw.net).

**Xingpeng Li** (S'13) received his B.S. and M.S. degree in electrical engineering in 2010 and 2013 from Shandong University and Zhejiang University, China, respectively. He is currently pursuing the Ph.D. degree in electrical engineering and the M.S. degree in industrial engineering at Arizona State University, Tempe, AZ, USA. His research interests include power system operations, planning, and optimization, microgrids, energy markets, and smart grid. He previously worked with ISO New England, Holyoke, MA, USA.

**Mostafa Sahraei-Ardakani** (M'06) received the PhD degree in energy engineering from Pennsylvania State University, University Park, PA in 2013. He also holds the B.S. and M.S. degrees in electrical engineering from University of Tehran, Iran. Currently, he is a post-doctoral scholar in the School Electrical, Computer, and Energy Engineering at Arizona State University. His research interests include energy economics and policy, electricity markets, power system optimization, transmission network modeling, and smart grids.

**Pranavamoorthy Balasubramanian** (S'08) received the B.E. degree in electrical and electronics engineering and M.E. degree in instrumentation engineering in 2009 and 2011 respectively from Anna University, Chennai, TN, India. He is currently working towards the Ph.D. degree in electrical engineering at Arizona State University. His research interests include power system operations and control, energy markets, smart grids, micro-grids, electrical machines, renewable energy sources, and process control.

**Mojdeh Abdi-Khorsand** (S'10) received the B.S. degree in electrical engineering from Mazandaran University, Babol, Iran, in 2007 and the M.S. degree in electrical engineering from the Iran University of Science and Technology, Tehran, Iran, in 2010. She is currently pursuing a Ph.D. degree and working as a research associate in the School of Electrical, Computer, and Energy Engineering at Arizona State University. Her research interests include power system operations, electricity markets, transmission switching, transient stability study and power system protection.

**Kory W. Hedman (S' 05, M' 10)** received the B.S. degree in electrical engineering and the B.S. degree in economics from the University of Washington, Seattle, in 2004 and the M.S. degree in economics and the M.S. degree in electrical engineering from Iowa State University, Ames, in 2006 and 2007, respectively. He received the M.S. and Ph.D. degrees in industrial engineering and operations research from the University of California, Berkeley in 2008 and 2010 respectively.

Currently, he is an assistant professor in the School of Electrical, Computer, and Energy Engineering at Arizona State University. He previously worked for the California ISO (CAISO), Folsom, CA, on transmission planning and he has worked with the Federal Energy Regulatory Commission (FERC), Washington, DC, on transmission switching. His research interests include power systems operations and planning, electricity markets, power systems economics, renewable energy, and operations research.

**Robin Podmore (M'73, F'96)** received the Bachelors and Ph.D. degrees in Electrical Engineering from University of Canterbury, N.Z in 1968 and 1973. In 1973 he worked as a Post-doctoral fellow at University of Saskatchewan, Saskatoon Canada. From 1974 to 1978 he managed the Power Systems Research group at Systems Control, Palo Alto, CA. From 1979 to July 1990 he was director and Vice President of Business Development with ESCA Corporation (now Alstom Grid), Bellevue WA. In July 1990 he founded and is President of Incremental Systems Corporation. He has been an industry leader and champion for open energy management systems, the Common Information Model, affordable and usable Operator Training Simulators and now affordable energy solutions for developing communities. He is a licensed Professional Engineer in the state of California. He is IEEE PES Vice President of New Initiatives and Outreach. Robin Podmore is a member of the USA National Academy of Engineering.